\documentstyle[12pt]{article}
\input tcilatex

\QQQ{Language}{
American English
}

\begin{document}

\begin{center}

{\Large La Place, Weimar, Schiller \\and the Birth of Black Hole Theory}

\vspace{6ex}

{\large H. Stephani}

\vspace{3ex}
{\small Theoretisch-Physikalisches Institut der Universit\"at Jena\\
ste@tpi.uni-jena.de}

\vspace{2ex}
{\scriptsize 17.4.03}

\vspace{8ex}

{\large {\bf Abstract}}

 \end{center}

From the original papers an interesting and surprising connection 
is established between La Place and his early discovery that black 
holes may occur in nature, and a poem by Schiller.

\newpage

\twocolumn
{\small 
\begin{center}

Allgemeine\\
Geographische\\ \vspace{1ex}
{\Large Ephemeriden.}\\
{\footnotesize Verfasset} \\ 
{\scriptsize von}\\ 
einer Gesellschaft Gelehrten\\ 
{\footnotesize und herausgegeben}\\ 
{\scriptsize von}\\
F. von Zach\\ \vspace{1ex} 
{\scriptsize H.S.G.
Obristwachtmeister und Director der herzoglichen \\ \vspace{-1ex}
Sternwarte Seeberg bey Gotha.}\\ \vspace{2ex}
Vierter Band.\\
\rule{6cm}{0.4mm}\\
Weimar\\ Im Verlage des Industrie-Comptoirs\\ 1799.\\
\vspace{3ex}

I.\\ABHANDLUNGEN.\\ 1.\\ Beweis\\des Satzes, da\ss\ die anziehende Kraft bey
einem Weltk\"orper so gro\ss\ seyn k\"onne, da\ss\ das Licht davon nicht
ausstr\"omen kann.\footnote{%
Diesen Satz, da\ss\ ein leuchtender K\"orper des Weltalls von gleicher
Dichtigkeit mit der Erde, dessen Durchmesser 250 mahl gr\"o\ss er w\"are,
als der der Sonne, verm\"oge seiner anziehenden Kraft keinen von seinen
Lichtstrahlen bis zu uns schicken k\"onne, da\ss\ folglich gerade die
gr\"o\ss ten K\"orper unseres Weltgeb\"audes uns unsichtbar bleiben
k\"onnen, hat La Place in seiner {\em Exposition du Systeme du Monde}
Part. II P. 305 ohne Beweis aufgestellt; hier ist er. Vgl.\ A. E. G. May 1798
S. 603 {\em v.\ Z.}}) \\Von\\Peter Simon La Place\\
\end{center}

{\bf 1}) Wenn $v$ die Geschwindigkeit, $t$ die Zeit und $s$ der w\"ahrend
dieser Zeit gleichf\"ormig durchlaufe-
\begin{center}
General\\
Geographical\\ \vspace{1ex}
{\Large Ephemeris.}\\
{\footnotesize Composed} \\ 
{\scriptsize by}\\ 
a Society of  Learned\\ 
{\footnotesize and brought out}\\ 
{\scriptsize by}\\
F. von Zach\\ \vspace{1ex} 
{\scriptsize H.S.G.
Major and Director of the Ducal\\ \vspace{-1ex}
Observatory Seeberg near Gotha.}\\ \vspace{2ex}
Fourth Volume.\\
\rule{6cm}{0.4mm}\\
Weimar\\ Published by the Industrie-Comptoirs\\ 1799.\\
\vspace{3ex}

I.\\PAPERS.\\ 1.\\ Proof\\ of the theorem that the attractive power of a 
heavenly body can be so large that light cannot leak out from it.\footnote{%
This theorem, that a luminous body in space, of equal density with the Earth,
whose diameter was 250 times that of the Sun, by virtue of its attractive
power is not able to send any of its beams of light as far as to us, and that
consequently just the largest bodies of our universe may be imperceptible, 
has been advanced by La Place in his {\em Exposition du Systeme du Monde}
Part. II P. 305 without proof; here it is. Cp.\ A. E. G. May 1798
S. 603 {\em v.\ Z.}\\}) \\By\\Peter Simon La Place\\
\end{center}

{\bf 1}) If $v$ is the velocity, $t$ the time and $s$ the interval uniformly 
passed during that time, then

\noindent nen Raum ist, so ist bekanntlich $v=s/t$.

{\bf 2}) Ist die Bewegung nicht gleichf\"ormig, so mu\ss\ man, um den Werth
von $v$ in jedem Augenblick zu haben, den in diesem Zeittheilchen d$t$
durchlaufenen Raum d$s$ ineinander dividiren, n\"amlich $v=$d$s/$d$t$; weil
die Geschwindigkeit in einem unendlich kleinen Zeittheilchen unver\"anderlich
und al\-so die Bewegung gleichf\"ormig angenommen werden kann.

{\bf 3}) Eine immerfort wirkende Kraft wird die Ge\-schwindigkeit zu \"andern
streben. Diese \"An\-derung der Geschwindigkeit, n\"amlich d$v$, ist das
nat\"urliche Ma\ss\ der Kraft. Da aber jede Kraft in doppelter Zeit doppelte
Wirkung hervorbringt, so mu\ss\ man noch die Aenderung der Geschwindigkeit
durch die Zeit d$t$, in welcher sie von der Kraft $P$ hervorgebracht wurde,
dividiren, und man wird da\-durch einen allgemeinen Ausdruck f\"ur die Kraft $%
P $ erhalten, n\"amlich $P=\,$d$v/$d$t=\,$d.(d$s/$d$t)/$d$t.$ Nun ist, wenn d%
$t $ best\"andig ist, d.(d$s/$d$t$)$\,=(\,$d.d$s)/$d$t,$ folglich $P=$dd$s/$d%
$t^2.$

{\bf 4}) Es sey die Attractions-Kraft eines K\"orpers = $M$; ein zweyter
K\"orper z.B. ein Lichttheilchen befindet sich in der Entfernung $r$; die
Wirkung der Kraft $M$ diese Lichttheilchens wird $-M/rr$ seyn; das Zeichen $%
-$ deswegen, weil die Wirkung von $M$ der Bewegung des Lichtes
entgegengesetzt ist.

{\bf 5}) Nun ist nach ({\bf 3}) diese Kraft auch $=\,$dd$r/$d$t^2,$ folglich 
$-M/rr=$dd$r/$d$t^2=-Mr^{-2},$ man multiplicire mit d$r$; d$r\,$dd$r/$d$%
t^2=-M $d$r\,r^{-2},$ integriert, $\tfrac 12$dtr$^2/$d$t^2=c+Mr^{-1},$ wo $c$
eine best\"andige Gr\"o\ss e ist, oder (d$r/$d$t$)$^2=2c+2Mr^{-1}.$ Nun ist
nach ({\bf 2}) d$r/$d$t=$ der Geschwindigkeit $v$ folglich $v^2=2c+2Mr^{-1}$
wo $v$ die Geschwindigleit des Lichttheilchens in der Entfernung $r$ ist.

{\bf 6}) Um nun die Constante $c$ zu bestimmen, sey $R$ der Halbmesser des
anziehenden K\"orpers, $a$ die Geschwindigkeit des Lichts in der Entfernung
$R$ folglich an der Oberfl\"ache des anziehenden K\"or-
\noindent everybody knows that $v=s/t$.

{\bf 2}) If the motion is not uniform, then, to get the value of
von $v$ at any instant, one has to divide the interval d$s$ passed 
in the time-particle  d$t$ into each other, that is  $v=$d$s/$d$t$; since
the velocity does not change in an infinitely small time-particle 
and so the motion may be assumed to be uniform.\\ \\

{\bf 3}) A continually acting force will tend to alter the velocity.
This change in the velocity, namely d$v$, is the natural measure 
of the force. Since in a twofold time any force will produce a 
twofold effect, one has still to divide the change of velocity by the time
d$t$, in which this change was brought forth by the force $P$, and 
by this one will obtain a general expression for the force, namely 
$P=\,$d$v/$d$t=\,$d.(d$s/$d$t)/$d$t.$ Then, if d%
$t $ is constant, d.(d$s/$d$t$)$\,=(\,$d.d$s)/$d$t,$ that is $P=$dd$s/$d%
$t^2.$\\ \\

{\bf 4}) Let  $M$ be the attractive force of a body; a second body, 
e.g.\ a light-particle, is at a distance $R$; the action of the 
force $M$ ((on)) this light-particle will be $-M/rr$; the sign $%
-$ since the action of $M$ is opposite to the motion 
of the light.\\ 

{\bf 5}) Now because of ({\bf 3}) this force also is $=\,$dd$r/$d$t^2,$ 
consequently $-M/rr=$dd$r/$d$t^2=-Mr^{-2},$
on multiplication with d$r$; d$r\,$dd$r/$d$%
t^2=\\-M $d$r\,r^{-2},$ and on integration, $\tfrac 12$dtr$^2/$d$t^2
=c+Mr^{-1},$ where $c$ is a constant, or (d$r/$d$t$)$^2=2c+2Mr^{-1}.$ 
Now because of ({\bf 2}), d$r/$d$t=$ the velocity $v$ consequently 
$v^2=2c+2Mr^{-1}$ where $v$ is the velocity of the light-particle 
at the distance $r$.

{\bf 6}) To determine now the constant $c$, let $R$ be the radius of 
the attracting body, $a$ the velocity of light at the distance $R$, 
hence at the surface of the  attracting body, then from ({\bf 5}) one obtains 
$a^2=$

\noindent pers, so erh\"alt man
aus ({\bf 5}) $a^2=2c+2M/R$ folglich $2c=a^2-2M/R$ die\ss\, in die vorige
Gleichung gesetzt, gibt $v^2=a^2-2M/R+2M/r.$

{\bf 7}) Eines andern anziehenden K\"orpers Halbmes\-ser sey $R^{\prime }$,
seine Attractionskraft sey $iM$, die Ge\-schwindigkeit des Lichts in der
Entfernung $r$ sey $v^{\prime }$ so ist verm\"oge der Gleichung 
in ({\bf 6})
$$
v^{\prime \,2}=a^2-\frac{2iM}{R^{\prime }}+\frac{2iM}r 
$$

{\bf 8}) Setzt man $r$ unendlich gro\ss\ , so verschwindet das letzte Glied
der vorhergehenden Gleichung und man erh\"alt%
$$
v^{\prime \,2}=a^2-\frac{2iM}{R^{\prime }}. 
$$
Die Entfernung der Fixsterne ist so gro\ss\ , da\ss\ man zu dieser Annahme
berechtigt ist.

{\bf 9}) Die anziehende Kraft des zweyten K\"orpers sey so gro\ss , da\ss\
das Licht nicht austr\"omen kann; dies l\"a\ss t sich analytisch am
bequemsten so aus\-dr\"ucken: die Geschwindigleit des Lichts $v^{\prime }$ ist
gleich Null. Diesen Werth von $v^{\prime }$ in der Gleichung f\"ur $%
v^{\prime }$ (8) gesetzt, wird eine Gleichung geben, aus der sich die Masse $%
iM$ wird herleiten lassen, bey welcher dieser Umstand Statt findet. Man hat
also $0=a^2-2iM/R^{\prime }$ oder $a^2=2iM/R^{\prime }$.

{\bf 10}) Um $a$ zu bestimmen, sey der erste anzie\-hende K\"orper die Sonne,
so wird $a$ die Geschwin\-digkeit des Sonnenlichts an der Oberfl\"ache der
Son\-ne seyn. Die anziehende Kraft der Sonne ist aber in Vergleichung mit der
Geschwindigkeit des Lichts so klein, da\ss\ man diese Geschwindigkeit als
gleichf\"ormig annehmen kann. Aus dem Ph\"anomen der Aberration erhellet,
da\ss\ die Erde 20''$\tfrac 14$ in ihrer Bahn durchl\"auft, w\"ahrend das
Licht von der Sonne bis zur Erde k\"ommt, folglich: es sey $V$ die mittlere
Geschwindigkeit der Erde in ihrer Bahn, so wird man haben \pagebreak

\noindent  $2c+2M/R$ therefore $2c=a^2-2M/R$ this, substituted 
into the forgoing equation, gives $v^2=a^2-2M/R+2M/r.$

{\bf 7}) Let $R^{\prime }$ be the radius of another attractive body, $iM$ 
its attractive force, and $v^{\prime }$ the velocity of light 
at the distance $r$, then by virtue of the equation in ({\bf 6})\vspace{-1ex}
$$
v^{\prime \,2}=a^2-\frac{2iM}{R^{\prime }}+\frac{2iM}r \vspace{-1ex}
$$
\\
{\bf 8}) If one takes $r$ infinitely large, then the last term of 
the forgoing equation vanishes, and one obtains
$$
v^{\prime \,2}=a^2-\frac{2iM}{R^{\prime }}. 
$$
The distance to the fixed stars is so large that 
this assumption is justified.

{\bf 9}) Assume now that the attractive power of the second body 
is so large that light cannot emanate; analytically this can best 
be expressed as:  the velocity $v^{\prime }$ of the light equals zero.
Substituting this value for $v^{\prime }$ into the equation {\bf 8})
for $v^{\prime }$ (8), will give an equation from which the mass
$iM$ can be derived for which this circumstance will take place.
Hence one has  $0=a^2-2iM/R^{\prime }$ or $a^2=2iM/R^{\prime }$.\\

{\bf 10}) To determine $a$, let the first acttractive 
body be the sun, then $a$ will be the velocity of the light of the sun 
at the surface of the sun. Compared to the velocity of the light, 
this attractive force will be so small that this velocity can be assumed 
to be constant. The phenomenon of aberration makes clear that the earth
travels 20''$\tfrac 14$ on its orbit while the light passes from the sun to 
the earth, therefore: if $V$ is the mean velocity of the earth on its orbit, 
one has \pagebreak  
$$
a:V=\text{radius}\footnote{%
in Secunden ausgedr\"uckt.}:20"\tfrac 14=1:\text{tang 
}20"\tfrac 14 
$$ 
{\bf 11}) Meiner Annahme in {\em Expos.\ du Syst.\ du Monde} Part II P. 205
gem\"a\ss , ist $R^{\prime }=250R.$ Nun verhalten sich Massen, wie die
Volumina der anzie\-henden K\"orper mit den Dichtigkeiten multiplicirt; die
Volumina wie die W\"urfel der Halbmesser, folglich die Massen, wie die
W\"urfel der Halbmesser mit den Dichtigkeiten multiplicirt. 
Es sey die Dich\-te
der Sonne $=1$, die des zweyten K\"orpers $=\rho $ so ist $M:1\,M=1\,R^3:\rho
\,R^{\prime 3}=iR^3:\rho 250R^3$ oder $i=250^3\rho .$

{\bf 12}) Man substituire die Werthe von $i$ und $R^{\prime }$ in die
Gleichung $a^2=2iM/R^{\prime }$, so erh\"alt man $a^2=2\,(250)^3\rho
M/250R=2\,(250)^2\rho M/R$ oder $\rho =a^2R/2\,(250)^2M$

{\bf 13}) Um $\rho $ zu haben, darf man nur noch $M$ bestimmen. Die Kraft
der Sonne $M$ ist in der Entfernung $D$ gleich $M/D^2.$ Es sey $D\,$ die
mittlere Entfernung der Erde, $V$ die mittlere Geschwindigkeit der Erde; so
ist diese Kraft auch gleich $V^2D$ (man sehe La Lande's Astronomie III \S\ 
3539) folglich\\ $M/D^2=V^2/D$ oder $M=V^2D$. Die\ss\ in die Glei\-chung 
f\"ur $%
\rho $ in ({\bf 12}) substituirt gibt \vspace{-1ex}
$$
\rho =\frac{a^2R}{2(250)^2V^2D}=\frac 8{(1000)^2}\left( \frac aV\right)
^2\left( \frac RD\right)  
$$
$$
\frac aV=\frac{\text{Geschw. d. Lichts}}{\text{Geschw. d. Erde}}=\frac 1{%
\text{tang }20"\tfrac 14}\text{ nach (10)} 
$$
$$\begin{array}{rl}
\dfrac RD=&\dfrac{\text{wahrem Halbmesser }\odot }{\text{ mittlern Entfernung }%
\odot }\\&\\
=&\text{tang. mittlern scheinbaren Halbmesser} \\
&\text{der }\odot . 
\end{array}$$
folglich $\rho =$8\ tang 16' 2''/(1000 tang $20"\tfrac 14)^2$ hieraus $\rho $
beynahe 4, oder so gro\ss , als die Dichte der Erde. \pagebreak
$$
a:V=\text{radius}\footnote{%
in seconds}:20"\tfrac 14=1:\text{tang 
}20"\tfrac 14 
$$
{\bf 11}) In agreement with my assumption in {\em Expos.\ du Syst.\ du Monde} 
Part II P. 205
one has $R^{\prime }=250R.$ Now the masses go as the volumes of the 
attracative bodies multiplied with the densities, and the volumes as the cubes
of the radii, therefore the masses as the cubes of the radii multiplied 
with the densities. Let the density of the sun be $=1$, that of the second 
body $=\rho $ then $M:1\,M=1\,R^3:\rho
R^{\prime 3}=i\,R^3:\rho \,250R^3$ or $i=250^3\rho .$\\

{\bf 12}) On substitution of the values of $i$ and $R^{\prime }$ 
into equation $a^2=2iM/R^{\prime }$, one obtains $a^2=2\,(250)^3\rho
M/250R=2\,(250)^2\rho M/R$ or $\rho =$\\ $ a^2R/2\,(250)^2M$

{\bf 13}) To have $\rho $, one now only has to determine $M$. 
In the distance $D$ the force of the sun equals $M/D^2.$ Let $D\,$ 
be the mean distance of the earth, $V$ the mean velocity of the earth;
then this force also equals is $V^2D$ (see also La Lande's Astronomie 
III \S\ 
3539) therefore $M/D^2=V^2/D$ or $M=V^2D$. This being substituted
into the equation for $%
\rho $ in ({\bf 12}) gives \vspace{-1ex}
$$
\rho =\frac{a^2R}{2(250)^2V^2D}=\frac 8{(1000)^2}\left( \frac aV\right)
^2\left( \frac RD\right)  
$$
$$
\frac aV=\frac{\text{velocity of the light}}{\text{velocity of the earth}}
=\frac 1{%
\text{tang }20"\tfrac 14}\text{ nach (10)} 
$$
$$\begin{array}{rl}
\dfrac RD=&\dfrac{\text{true radius} \odot }{\text{mean distance}%
 \odot }\\&\\ 
=&\text{tang. mean apparent radius} \\ 
&\text{der }\odot . 
\end{array}$$
hence  $\rho =$8 tang 16' 2''/(1000 tang $20"\tfrac 14)^2$ From this $\rho $
nearly 4, or as large as the density of the earth.\\
\vspace{1ex}
\begin{centering} (Translation anonymous, 2003)\end{centering}

\onecolumn

\subsection*{Peter Simon La Place and the Schwarzschild Black hole}
One of the first proofs that there may be black holes in our universe is
due to La Place. In his paper with the title ''Proof of the theorem that a
heavenly body can be such large that light cannot leak out from it'' he uses
Newtonian gravity theory to calculate the escape velocity of a star, and
shows that it might be larger than that of light.

When reading his paper, one might be surprised to see that it is written in
German and even more surprised that it was published in Weimar, in a journal
with the title ''Allgemeine Geographische Ephemeriden'' 
(for the benefit of the
readers who may not have access to this paper we gave it in full). 
Why in German, and why exactly in Weimar, the town famous due 
to Goethe and Schiller? And who else dealt
there with ''Schwerkraft'' (gravity) - the German word La Place uses - 
in those days? Today, to get answers to this kind of questions, one is inclined 
to consult a database
or the internet. But for this old stuff other sources have to be found.

The most famous source for German words and notions is Grimm's
"Deutsches W\"orterbuch''. In this many-volume dictionary all
German words found in the literature of the three centuries from Luther to
Goethe have been collected, a task, the Grimm brothers could not finish
themselves, it took them and their successors more than 100 years. If here
we look up the word ''Schwerkraft'', we learn that the first ever to use
this German word (instead of the Latin-based gravitation) was Friedrich von
Schiller, in a poem called ''The Fountain''.

This leads us back to Weimar, and one may ask whether there is more what
Schiller and La Place have in common than only the Schwerkraft - one coining
the word, and the other using its physics - and the place where they worked
and published. Here the German philosopher Kant enters. Using Kant's ideas,
La Place developed a model of the origin of our planetary system, called the
Kant-Laplace nebular hypothesis. And Kant had also a big influence on
Schillers thinking, so big and enduring that Goethe and others were
concerned that this influence might seriously damage Schiller's poetic
abilities.

Putting all these pieces together one may wonder whether there are 
more papers by Schiller concerned with gravity.

\newpage
\twocolumn

\begin{center}{\bf {\large Der Taucher}}\end{center}
\vspace{2ex}
{\footnotesize

\noindent{\bf 1} $,\!\!,$Wer wagt es, Rittersmann oder Knapp', \\
Zu tauchen in diesen Schlund? \\ 
Einen goldnen Becher werf ich hinab, \\
Verschlungen schon hat ihn der schwarze Mund.\\
Wer mir den Becher kann wieder zeigen, \\
Er mag ihn behalten, er ist sein eigen."\\

\noindent{\bf  2} Der K\"onig spricht es und wirft von der H\"oh' \\
Der Klippe, die schroff und steil \\
Hinaush\"angt in die unendliche See, \\
Den Becher in der Charybde Geheul. \\
$,\!\!,$Wer ist der Beherzte, ich frage
wieder, \\
Zu tauchen in diese Tiefe nieder?'' \\

\noindent{\bf  3} Und die Ritter, die Knappen um
ihn her \\
Vernehmen's und schweigen still, \\
Sehen hinab in das wilde Meer, \\
Und Keiner den Becher gewinnen will. \\
Und der K\"onig zum dritten Mal wieder
fraget:\\
$,\!\!,$Ist Keiner, der sich hinunter waget?'' \\ 

\noindent{\bf  4} Doch Alles noch stumm
bleibt wie zuvor, \\
Und ein Edelknecht, sanft und keck, \\
 Tritt aus der
Knappen zagendem Chor, \\
Und den G\"urtel wirft er, den Mantel weg, \\
Und alle
die M\"anner umher und Frauen \\
Auf den herrlichen J\"ungling verwundert  \noindent {\bf  } 
schauen.\\ 

\noindent{\bf 5 } 
Und wie er tritt an des Felsen Hang \\
Und blickt in den Schlund
hinab,\\
Die Wasser, die sie hinunterschlang,\\ 
Die Charybde jetzt br\"ullend
wiedergab, \\
Und wie mit des fernen Donners Getose\\ 
Entst\"urzen sie
sch\"aumend dem finstern Scho\ss e. \\

\noindent {\bf 6} Und es wallet und siedet und brauset
und zischt, \\
Wie wenn Wasser mit Feuer sich mengt, \\
Bis zum Himmel spritzet
der dampfende Gischt, \\
Und Flut auf Flut sich ohn' Ende dr\"angt, \\
Und will
sich nimmer ersch\"opfen und leeren,\\ 
Als wollte das Meer noch ein Meer
geb\"aren. \\

\begin{center}{\bf {\large The Diver}}\end{center}
\vspace{2ex}

\noindent ''What knight or what vassal will be so bold\\ 
As to plunge in the gulf below?\\ 
See! I hurl in its depths a goblet of gold,\\
Already the waters over it flow.\\ 
The man who can bring back the goblet tome,\\ 
May keep it henceforward, - his own it shall be.'' \\

\noindent Thus speaks the king, and he hurls from the height 
\\Of the cliffs that, rugged and steep, \\Hang over the boundless sea, with
strong might, \\The goblet afar, in the bellowing deep. \\''And who'll be so
daring, - I ask it once more, \\As to plunge in these billows that wildly
roar?'' \\

\noindent And the vassals and knights of high degree\\ 
Hear
his words, but silent remain.\\ 
They cast their eyes on the raging sea,\\
And none will attempt the goblet to gain.\\ 
And a third time the question is
asked by the king:\\
 ''Is there none that will dare in the gulf now to
spring?''\\ 

\noindent Yet all as before in silence stand,\\ 
When a page,with a modest pride,\\ 
Steps out of the timorous squirely band,\\ 
And his girdle and mantle soon throws aside,\\ 
And all the knights, and the ladies too,\\ 
The noble stripling with wonderment view.\\ 

\noindent And when he draws nigh to the rocky brow,\\ 
And looks in the gulf so black,\\ 
The waters that she had swallowed but now,\\
The howling Charybdis is giving back;\\ 
And, with the distant thunder's dull sound\\ 
From her gloomy womb they all-foaming rebound\\

\noindent And it boils and it roars, and it hisses and
seethes.\\ 
As when water and fire first blend;\\ 
To the sky spurts the foam
in steam-laden wreaths,\\ 
And wave presses hard upon wave without end.\\ 
And the ocean will never exhausted be,\\ 
As if striving to bring forth another sea.\\

\noindent {\bf 7} Doch endlich, da legt sich die wilde Gewalt, \\
Und schwarz aus dem
wei\ss en Schaum \\
Klafft hinunter ein g\"ahnender Spalt,\\ 
Grundlos, als ging's
in den H\"ollenraum,\\ 
Und rei\ss end sieht man die brandenden Wogen \\
Hinab
in den strudelnden Trichter gezogen. \\

\noindent {\bf 8} Jetzt schnell, eh' die Brandung
wiederkehrt, \\
Der J\"ungling sich Gott befiehlt,\\ 
Und - ein Schrei des
Entsetzens wird rings geh\"ort, \\
Und schon hat ihn der Wirbel
hinweggesp\"ult, \\
Und geheimnisvoll \"uber dem k\"uhnen Schwimmer \\
Schlie\ss t
sich der Rachen, er zeigt sich nimmer. \\

\noindent {\bf 9} Und stille wird's \"uber dem
Wasserschlund, \\
In der Tiefe nur brauset es hohl, \\
Und bebend h\"ort man
von Mund zu Mund: \\
$,\!\!,$Hochherziger J\"ungling, fahre wohl!'' \\
Und hohler und
hohler h\"ort man's heulen,\\ 
Und es harrt noch mit bangem, mit schrecklichem
Weilen. \\

\noindent {\bf 10} Und, w\"urfst du die Krone selber hinein\\ 
Und spr\"achst: Wer mir
bringet die Kron', \\
Er soll sie tragen und K\"onig seyn!\\ 
Mich gel\"ustete
nicht nach dem theuren Lohn.\\ 
Was die heulende Tiefe da unter verhehle, \\
Das
erz\"ahlt keine lebende gl\"uckliche Seele. \\

\noindent {\bf 11} Wohl manches Fahrzeug, vom
Strudel gefa\ss t,\\ 
Scho\ss  j\"ah in die Tiefe hinab, \\
Doch zerschmettert nur
rangen sich Kiel und Mast,\\
 Hervor aus dem Alles verschlingenden Grab - \\
Und
heller und heller, wie Sturmes Sausen, \\
H\"ort man's n\"aher und immer
n\"aher brausen.\\ 

\noindent {\bf 12} Und es wallet und siedet und brauset und zischt, \\
Wie wenn
Wasser mit Feuer sich mengt, \\
Bis zum Himmel spritzet der dampfende Gischt,\\
Und Well' auf Well' sich ohn' Ende dr\"angt, \\
Und wie mit des fernen Donners
Getose \\
Entst\"urzt es br\"ullend dem finstern Scho\ss e. \\

\vspace{6ex}
\noindent But at length the wild tumult seems pacified,\\ And
blackly amid the white swell\\ A gaping chasm its jaws opens wide,\\ As if
leading down to the depths of hell:\\ And the howling billows are seen by
each eye\\ Down the whirling funnel all madly to fly\\ 

\noindent Then quickly, before the breakers rebound,\\ The
stripling commends him to Heaven,\\ And - a scream of horror is heard
around, -\\ And now by the whirlpool away he is driven,\\ And secretly over
the swimmer brave\\ Close the jaws, and he vanishes 'neath the dark wave.\\ 

\noindent O'er the watery gulf dread silence now lies,\\ But
the deep sends up a dull yell,\\ And from mouth to mouth thus trembling it
flies:\\ ''Courageous stripling, oh, fare thee well!''\\ And duller and
duller the howls recommence,\\ While they pause in anxious and fearful
suspense.\\ 

\noindent ''If even thy crown in the gulf thou shouldst fling,%
\\ And shouldst say, 'He who brings it to me\\ Shall wear it henceforward,
and be the king,'\\ Thou couldst tempt me not e'en with that precious fee;\\
What under the howling deep is concealed\\ To no happy living soul is
revealed! ''\\ 

\noindent Full many a ship, by the whirlpool held fast, \\%
Shoots straightway beneath the mad wave,\\
And, dashed to pieces, the hull and
the mast\\ Emerge from the all-devouring gave. -\\ And the roaring
approaches still nearer and nearer,\\ Like the howl of the tempest, still
clearer and clearer.\\ 

\noindent And it boils and it roars, and it hisses and
seethes,\\ As when water and fire first blend;\\ To the sky spurts the foam
in steam-laden wreaths \\And wave passes hard upon wave without end,\\ And,
with the distant thunder's dull sound,?\\ From the ocean-womb they
all-bellowing bound.\\ 
\vspace{6ex}

\noindent {\bf 13} Und sieh'! aus dem
finster flutenden Scho\ss, \\
Da hebet sich's schwanenwei\ss , \\
Und ein Arm und
ein gl\"anzender Nacken wird blo\ss ,\\ 
Und es rudert mit Kraft und mit
emsigem Flei\ss , \\
Und er ist's, und hoch in seiner Linken \\
Schwingt er den
Becher mit freudigem Winken.\\ 

\noindent {\bf 14} Und athmete lang und athmete tief \\
Und begr\"u\ss te das himmlische Licht.\\ 
Mit Frohlocken es Einer dem Andern rief: \\
$,\!\!,$Er lebt! Er ist da! Es behielt ihn nicht! \\
Aus dem Grab, aus der strudelnden
Wasserh\"ohle \\
Hat der Brave gerettet die lebende Seele.'' \\

\noindent {\bf 15} Und er kommt, es
umringt ihn die jubelnde Schar, \\
Zu des K\"onigs F\"u\ss en er sinkt, \\
Den
Becher reicht er ihm kniend dar,\\ 
Und der K\"onig der lieblichen Tochter
winkt,\\ 
Die f\"ullt ihn mit funkelndem Wein bis zum Rande, \\
Und der J\"ungling
sich also zum K\"onig wandte:\\ \\

\noindent {\bf 16}  $,\!\!,$Lange lebe der K\"onig! Es freue sich,\\
Wer da atmet im rosigten Licht! \\
Da unten aber ist's f\"urchterlich, \\
Und der
Mensch versuche die G\"otter nicht \\
Und begehre nimmer und nimmer zu schauen,\\
Was sie gn\"adig bedeckten mit Nacht und Grauen."\\ 

\noindent {\bf 17} $,\!\!,$Es ri\ss  mich hinunter
blitzesschnell,  \\
Da st\"urzt' mir aus felsigem Schacht\\ 
Wildflutend entgegen
ein rei\ss ender Quell;\\ 
Mich packte des Doppelstroms w\"uthende Macht, \\
Und, wie einen Kreisel, mit schwindendelm  Drehen \\
Trieb mich's um, ich konnte nicht widerstehen." \\

\noindent {\bf 18} $,\!\!,$Da zeigte mir Gott, zu dem ich rief,\\
In der h\"ochsten schrecklichen Noth, \\
Aus der Tiefe ragend, ein Felsenriff,\\ 
Das erfa\ss t' ich behend und entrann dem Tod. \\
Und da hing auch der Becher an spitzen
Korallen,\\
Sonst w\"ar er ins Bodenlose gefallen."

\vspace{6ex}

\noindent And lo! from the darkly flowing tide\\ Comes a
vision white as a swan,\\ And an arm and a glistening neck are descried,\\
With might and with active zeal steering on;\\ And 'tis he, and behold! his
left hand on high\\ Waves the goblet, while beaming with joy is his eye.\\ 

\noindent Then breathes he deeply, then breathes he long,\\
And blesses the light of the day;\\ While gladly exclaim to each other the
throng:\\ ''He lives! he is here! he is not the sea's prey!\\ From the tomb,
from the eddying waters' control,\\ The brave one has rescued his living
soul!''\\ 

\noindent And he comes, and they joyously round him stand;\\
At the feet of the monarch he falls, -\\ The goblet he, kneeling, puts in
his hand,\\ And the king to his beauteous daughter calls,\\ Who fills it
with sparkling wine to the brim;\\ The youth turns to the monarch, and
speaks thus to him:\\ 

\noindent ''Long life to the king! Let all those be glad\\
Who breathe in the light of the sky!\\ For below all is fearful, of moment
sad;\\ Let not man to tempt the immortals e'er try,\\ Let him never desire
the thing to see\\ That with terror and night they veil graciously.\\ 

\noindent ''I was torn below with the speed of light,\\ When
out of a cavern of rock\\ Rushed towards me a spring with furious might;\\ I
was seized by the twofold torrent's wild shock,\\ And like a top, with a
whirl and a bound,\\ Despite all resistance, was whirled around.\\ 

\noindent ''Then God pointed out, - for to Him I cried\\ In
that terrible moment of need, -\\ A craggy reef in the gulf's dark side;\\ I
seized it in haste, and from death was then freed.\\ And there, on sharp
corals, was hanging the cup, -\\ The fathomless pit had else swallowed it up.%
\\ 
\vspace{4ex}

\noindent {\bf 19} $,\!\!,$Denn unter mir lag's noch
bergetief,\\ 
In purpurner Finsterni\ss  da, \\
Und, ob's hier dem Ohre gleich
ewig schlief, \\
Das Auge mit Schaudern hinuntersah,\\ 
Wie's von Salamandern und Molchen und Drachen \\
Sich regt' in dem furchtbaren H\"ollenrachen."\\

\noindent {\bf 20}  $,\!\!,$Schwarz
wimmelten da, in grausem Gemisch,\\ 
Zu scheuslichen Klumpen geballt,\\
Der
stachligte Roche, der Klippenfisch, \\
Des Hammers gr\"auliche Ungestalt, \\
Und dr\"auend wies mir die grimmigen Z\"ahne\\ 
Der entsetzliche Hay, des Meeres
Hy\"ane."\\ 

\noindent {\bf 21} $,\!\!,$Und da hing ich und 
war's mir mit Grausen bewu\ss t, \\
Von der menschlichen H\"ulfe so weit,\\ 
Unter Larven die einzige f\"uhlende Brust,\\
Allein in der gr\"a\ss lichen Einsamkeit, \\
Tief unter dem Schall der
menschlichen Rede\\ 
Bei den Ungeheuern der traurigen Oede." \\ \\

\noindent {\bf 22} $,\!\!,$Und schaudernd
dacht ich's - da kroch's heran,\\
Regte hundert Gelenke zugleich,\\ 
Will schnappen nach mir; in des Schreckens Wahn \\
La\ss  ich los der Koralle umklammerten Zweig,\\ 
Gleich fa\ss t mich der Strudel mit rasendem Toben,\\
Doch es war mir zum Heil, er ri\ss  mich nach Oben.'' \\ \\

\noindent {\bf 23} Der K\"onig darob sich
verwundert schier\\
Und spricht: $,\!\!,$Der Becher ist dein, \\
Und diesen Ring noch
bestimm' ich dir, \\
Geschm\"uckt mit dem k\"ostlichsten Edelgestein,\\
Versucht du's noch einmal und bringst mir Kunde, \\
Was du sahst auf des Meeres
tiefunterstem Grunde.'' \\

\noindent {\bf 24} Das h\"orte die Tochter mit weichem Gef\"uhl,\\ 
Und
mit schmeichelndem Munde sie fleht: \\
$,\!\!,$La\ss t, Vater, genug seyn das
grausame Spiel! \\
Er hat Euch bestanden, was Keiner besteht, \\
Und, k\"onnt Ihr
des Herzens Gel\"uste nicht z\"ahmen, \\
So m\"ogen die Ritter den Knappen
besch\"amen.'' \\

\noindent ''For under me lay it, still mountain-deep,\\ In a
darkness of purple-tinged dye,\\ And though to the ear all might seem then
asleep\\ With shuddering awe 'twas seen by the eye\\ How the salamanders'
and dragons' dread forms\\ Filled those terrible jaws of hell with their
swarms.\\ 

\noindent ''There crowded, in union fearful and black,\\ In a
horrible mass entwined,\\ The rock-fish, the ray with the thorny back,\\ And
the hammer-fish's misshapen kind,\\ And the shark, the hyena dread of the
sea,\\ With his angry teeth, grinned fiercely on me.\\ 

\noindent ''There hung I, by fulness of terror possessed,\\
Where all human aid was unknown,\\ Amongst phantoms, the only sensitive
breast,\\ In that fearful solitude all alone,\\ Where the voice of mankind
could not reach to mine ear, '\\ Mid the monsters foul of that wilderness
drear.\\ 

\noindent ''Thus shuddering methought - when a something\\
crawled near,\\ And a hundred limbs it out-flung,\\ And at me it snapped; -
in my mortal fear,\\ I left hold of the coral to which I had clung;\\ Then
the whirlpool seized on me with maddened roar,\\ Yet 'twas well, for it
brought me to light once more.''\\ 

\noindent The story in wonderment hears the king,\\ And he
says, ''The cup is thine own,\\ And I purpose also to give thee this ring,\\
Adorned with a costly , a priceless stone,\\ If thou'lt try once again, and
bring word to\\ me What thou saw'st in the nethermost depths of sea.''\\ 

\noindent His daughter hears this with emotions soft,\\ And
with flattering accent prays she: ''\\ That fearful sport, father, attempt
not too oft!\\ What none other would dare, he hath ventured for thee;\\ If
thy heart's wild longings thou canst not tame,\\ Let the knights, if they
can, put the squire to shame.''\\

\pagebreak

\noindent {\bf 25} 
Drauf der K\"onig greift nach dem Becher schnell, \\
In den Strudel ihn schleudert hinein:\\
$,\!\!,$Und schaffst du den Becher mir wieder zur
Stell', \\
So sollst du der trefflichste Ritter mir sein \\
Und sollst sie als
Ehgemahl heut' noch umarmen, \\
Die jetzt f\"ur dich bittet mit zartem
Erbarmen.'' \\ \\

\noindent {\bf 26} Da ergreift's ihm die Seele mit Himmelsgewalt, \\
Und es blitzt
aus den Augen ihm k\"uhn, \\
Und er siehet err\"othen die sch\"one Gestalt 
\\Und
sieht sie erbleichen und sinken hin -\\
 Da treibt's ihn, den k\"ostlichen
Preis zu erwerben, \\
Und st\"urzt hinunter auf Leben und Sterben.\\

\noindent {\bf 27} Wohl
h\"ort man die Brandung, wohl kehrt sie zur\"uck, \\
Sie verk\"undigt der
donnernde Schall, \\
Da b\"uckt sich's hinunter mit liebendem Blick -\\
Es kommen, es kommen die Wasser all, \\
Sie rauschen herauf, sie rauschen nieder -\\
Den J\"ungling bringt keines wieder.\\

\vspace{4ex}

\begin{centering} (1797) \end{centering}

\pagebreak

\noindent The king then seizes the goblet in haste,\\ In the
gulf he hurls it with might:\\ ''When the goblet once more in my hands thou
hast placed,\\ Thou shalt rank at my court as the noblest knight,\\ And her
as a bride thou shalt clasp e'en today\\ Who for thee with tender compassion
cloth pray.''\\ 

\noindent Then a force, as from Heaven, descends on him there,%
\\ And lightning gleams in his eye,\\ And blushes he sees on her features so
fair,\\ And be sees her turn pale, and swooning lie;\\ Then eager the
precious guerdon to win,\\ For life or for death, lo! he plunges him in!\\ 

\noindent The breakers they hear, and the breakers return,\\
Proclaimed by a thundering sound;\\ They bend o'er the gulf with glances
that yearn\\ And the waters are pouring in fast around;\\ Though upwards and
downwards they rush and they rave,\\ The youth is brought back by no kindly
wave. 
\vspace{4ex}

\begin{centering}(Translation anonymous, 1902) \end{centering}

}
\onecolumn

\subsection*{What Johann Christoph Friedrich von Schiller has to say}

When searching Schiller's collected works for contributions to gravity, one
of course has to take into account that Schiller was a poet, not a
scientist. He will wrap up everything by a story on love, young and pretty
girls and brave boys, and instead of dealing with the depth, force and
loneliness of space, he will describe those of the sea or the human soul.
Keeping that in mind when searching Schiller's collected papers 
(fortunately the poems usually are in the first volume of such collections) 
we quite easily find what we are looking for : it is
his famous poem ''The Diver''. It describes in some detail, and rather 
beautifully, the encounter of men with a rotating black hole.

Once said, all pieces fit together, and we
shall not bore the reader by explaining everything lengthy and 
in much detail. Rather we shall give some hints, and ask the reader 
to enjoy the full text of
this poem. (Numbers refer to the relevant verses, and sometimes we 
refer to the original German wording when the translation does not 
reflect the meaning correctly.)

The opening is blunt and direct: The king looks over the infinite space
(the boundless sea), sees the black hole ({\bf 1}, the "Schwarze Schlund" = 
the black gorge) and sets the task of exploring it and recovering 
energy (gold). It is a rotating black hole ({\bf 7}, whirling funnel), 
which also ejects matter, but mainly swallows it via its 
accretion disc ({\bf 5}, {\bf 11}). The page sets out to explore the black 
hole, and happily returns. He tells that he was forced to corotate {\bf 17} 
(not with "the speed of light" - this a poetic translation 
of "blitzesschnell" = as fast as lightning, 
so don't blame Schiller for that!),
but luckily enough the page found that he could retrieve energy (the goblet) 
and was able to return: he had passed the surface of stationarity and been in the 
ergosphere. He also became aware of the cosmic censorship 
({\bf 15}, "... what the immortals veil graciously ..."). But 
the king wants to know what is behind 
the horizon; as politicians are, he does not listen to reasoning 
and sends the page a 
second time - right into the black hole and the death.

All this is a surprisingly exact description of a Kerr Black Hole, 
with a detailed poetic picture of all the debris 
in the accretion disc.

\subsection*{PS}
Let use end with the remark that certainly this early blossoming of gravity research 
had a strong influence on Jena, the town situated so near to 
Weimar and also home to Schiller before he went to Weimar, and made 
the people there
susceptible for the ideas of gravitation until our days.

\begin{center}
*
\end{center}
It has already been observed by psychoanalysts that the emotional story of the 
diver and the terrors of the sea which frighten men is
deeply anchored in what mankind loves and fears - this of course is also true 
for the black holes and may explain the attraction they find by the general 
public.
\begin{center}
*
\end{center}
\end{document}